\documentclass[prb,showpacs,twocolumn,aps,superscriptaddress,a4paper]{revtex4-1}
\usepackage{dcolumn,amssymb,amsmath,amsfonts,graphicx,latexsym,color,braket,amssymb,mathrsfs}

\begin{document}

\title{Density matrix of chaotic quantum systems}
\author{Xinxin Yang}
\affiliation{Department of Physics, Zhejiang Normal University, Jinhua 321004, People's Republic of China}
\author{Pei Wang}
\email{wangpei@zjnu.cn}
\affiliation{Department of Physics, Zhejiang Normal University, Jinhua 321004, People's Republic of China}
\date{\today}

\begin{abstract}
The nonequilibrium dynamics in chaotic quantum systems
denies a fully understanding up to now, even if thermalization in the long-time
asymptotic state has been explained by the eigenstate thermalization hypothesis which assumes
a universal form of the observable matrix elements in the eigenbasis of Hamiltonian.
It was recently proposed that the density matrix elements have also a universal form, which can be used to
understand the nonequilibrium dynamics at the whole time scale, from the transient regime to
the long-time steady limit. In this paper, we numerically test these assumptions for density matrix in the models of spins.
\end{abstract}

\maketitle

\section{\label{sec:level 1}Introduction}

The nonequilibrium dynamics of quantum many-body systems keeps on
attracting attention of both experimentalists~\cite{Bloch08}
and theorists~\cite{Polkovnikov11,Eisert15}. For integrable
systems, the case-by-case study of exact solutions revealed exotic properties
of quantum states driven out of equilibrium. The long-time asymptotic state is far from thermal
equilibrium, but should be described by the generalized Gibbs ensemble~\cite{rigol2007relaxation}.
On the other hand, for systems whose classical counterparts are
chaotic, it is widely believed that they thermalize in the
long time limit~\cite{rigol2008thermalization}. But the dynamics in the transient and intermediate time scale
is still hard to explore, due to the lack of a reliable analytical or numerical method.

The study of dynamics in quantum chaotic systems dated back to
the early days of quantum mechanics, when the question has been raised as to
how the statistical properties of equilibrium ensembles arise from
the unitary evolution of a pure quantum state~\cite{Neumann29,Goldstein10}.
A breakthrough was made in 1950s by Wigner~\cite{Wigner55,Wigner57,Wigner58}, who stated
that the statistics of the eigenenergies of a chaotic system should be as same as that of a random matrix
whose level spacing follows the Wigner-Dyson distribution.
This statement was verified by both experiments and numerical
simulations~\cite{rosenzweig1960repulsion,brody1981random,
bohigas1984characterization,schroeder1987statistical,guhr1998random}.
On the other hand, the level spacing
satisfies a Poisson distribution in an integrable system, according to Berry and Tabor~\cite{Berry77}.

In the random matrix theory (RMT), the eigenstates of Hamiltonian
are considered to be random vectors in the Hilbert space.
This oversimplified picture ignores the dependence of
eigenstate on the eigenenergy, and then fails to explain why the observable
is a function of energy or temperature of the system.
The eigenstate thermalization hypothesis (ETH)~\cite{deutsch1991quantum,
srednicki1994chaos,srednicki1999approach} overcomes this problem by assuming a generic form
of the matrix elements of observable operators in the eigenbasis of Hamiltonian:
\begin{equation}\label{eth}
O_{\alpha\beta}={O}\left(\bar{E}\right)\delta_{\alpha\beta}
+D^{-\frac{1}{2}}\left(\bar{E}\right)f_O\left(\omega,\bar{E}\right)R^O_{\alpha\beta},
\end{equation}
where $\alpha$ and $\beta$ are the eigenstates of Hamiltonian with
$E_{\alpha}$ and $E_{\beta}$ being their eigenenergies, respectively.
$\bar{E}= (E_{\alpha}+E_{\beta})/2$ and $\omega=E_{\alpha}-E_{\beta}$
denote the average energy and energy difference of $\alpha$ and $\beta$, respectively.
$D\left(\bar{E}\right)$ is the density of many-body states, which increases exponentially
with the system size.
${O}\left(\bar{E}\right)$ and $f_O\left(\omega,\bar{E}\right)$ are both smooth functions,
with the former describing how the expectation value of observable changes with energy.
The randomness of the eigenstates manifests itself in the random number
$R^O_{\alpha\beta}$, which has zero mean and unit variance according to definition.

When a chaotic system is driven out of equilibrium, its density matrix evolves according
to the quantum Liouville equation. In the asymptotic
long-time state, the off-diagonal elements of the density matrix obtain completely randomized phases,
therefore, only the diagonal elements have a contribution to the expectation value of observables. 
ETH builds the equivalence between the microcanonical
ensemble and diagonal ensemble, and then explains thermalization successfully.
Its correctness has been verified in plenty of numerical simulations~\cite{d2016quantum},
while its limitation was also noticed. ETH has to be modified for the order parameter in the presence of
spontaneous symmetry breaking~\cite{PhysRevE.92.040103}, and
it fails in a many-body localized system~\cite{Gornyi05,basko2006metal} which cannot thermalize.

In spite of the success of ETH, it cannot explain how an observable relaxes towards the steady value,
because it says nothing about the off-diagonal elements
of density matrix which are important in the transient and intermediate time scale.
The off-diagonal elements of density matrix are even the key for describing an asymptotic long-time state,
in the case that the thermodynamic limit and long-time limit
are noncommutative. This noncommutativity defines an important class of nonequilibrium states
- the nonequilibrium steady states in the study of mesoscopic transport.
A stationary current flows through a central region which is
coupled to multiple thermal reservoirs at different temperatures and chemical potentials.
The description of such a state goes beyond the ability of diagonal ensemble,
Gibbs ensemble or generalized Gibbs ensemble,
but requires the knowledge of off-diagonal elements.
This motivated one of the authors to propose the nonequilibrium
steady state hypothesis~(NESSH)~\cite{wang2017theory}.

The purpose of this paper is to test the hypotheses on density matrix in
the spin lattice models. We numerically diagonalize the Hamiltonians of
disordered XXZ model and two-dimensional Ising model.
We find that the off-diagonal elements are well described by NESSH.
At the same time, the diagonal element is a Gaussian function
of energy, which coincides with previous studies in different models~\cite{PhysRevLett.107.040601,
PhysRevE.56.5144}.

The rest of paper is organized as follows.
The assumptions about density matrix are explained in Sec.~\ref{sec:level 2}.
In Sec.~\ref{sec:level 3}, we define the spin models and show the numerical
results about diagonal elements. The off-diagonal elements are
discussed in Sec~\ref{sec:level 4}. Sec.~\ref{sec:level 6} summarizes our results.

\section{\label{sec:level 2}Assumptions about density matrix}

\subsection{Density matrix of a generic initial state}

Let us consider the real-time dynamics in an isolated system whose Hamiltonian is $\hat H$.
Without loss of generality, we suppose $t=0$ to be the initial time. At $t=0$,
the system is prepared in a pure quantum state $\ket{s}$.
According to quantum mechanics, the time-dependent expectation value of
an arbitrary observable can be expressed as
\begin{equation}\label{quench}
O(t)=\sum_{\alpha,\beta} e^{-i\omega t}\rho_{\alpha\beta}O_{\beta\alpha},
\end{equation}
where $\alpha$ and ${\beta}$ denote the eigenstates of $\hat H$,
and $\omega=E_{\alpha}-E_{\beta}$ is the difference between their eigenenergies.
$\rho_{\alpha\beta}=\bra{\alpha}\hat{\rho}\ket{\beta}$ with $\hat \rho =\ket{s}\bra{s}$
denotes the element of initial density matrix in the eigenbasis of $\hat H$,
and $O_{\beta\alpha}=\bra{\beta}\hat{O}\ket{\alpha}$ denotes the matrix element of
observable operator.

If the initial state is not an eigenstate of $\hat H$, the system is out of equilibrium at $t>0$.
To calculate $O(t)$, we need to know the eigenenergies, initial density matrix and observable matrix.
In the integrable systems, $E_\alpha$, $\rho_{\alpha\beta}$ and $O_{\alpha\beta}$
differ from model to model, and no common formulas for them are known up to now.
On the other hand, the chaotic systems are "similar"
to each other. RMT states that their eigenenergies all
follow the Wigner-Dyson distribution~\cite{kravtsov2012random}
\begin{equation}\label{WDdis}
P(E_1,E_2,\cdots) = \frac{1}{\mathcal{N}} \displaystyle e^{-\frac{E_1^2+E_2^2+\cdots}{2\sigma^2}}
\left| \prod_{\alpha>\beta}\left(E_\alpha-E_\beta\right) \right|,
\end{equation}
where $\sigma$ is connected to the energy bandwidth and $\mathcal{N}$
is a normalization constant. And according to ETH, $O_{\alpha\beta}$ has the universal
form~(\ref{eth}), independent of whether the system is made of fermions, bosons or spins
and in which dimensions. Once if we know the expression of $\rho_{\alpha\beta}$, $O(t)$ can be calculated.
Different from integrable models, our knowledge of $E_\alpha$ and $O_{\alpha\beta}$
in chaotic models is not precise, but only statistical. Eq.~(\ref{WDdis})
only gives the probability density of the eigenenergies, and Eq.~(\ref{eth}) contains a
random number $R^O_{\alpha\beta}$. We then expect that
$\rho_{\alpha\beta}$ should be expressed in a similar way.

Before discussing the form of $\rho_{\alpha\beta}$, we need to clarify
which kind of initial states are interesting to us. The initial state $\ket{s}$
should be some quantum state that we can prepare in laboratory.
Preparing a quantum state is usually equivalent to measuring
the state which inevitably causes the wave function collapsing
into an eigenstate of the observable operator. Therefore, it is natural
to choose $\ket{s}$ to be the common eigenstate of a set of observable operators.
For example, in a spin lattice model, we can choose $\ket{s}$ to be
the spin eigenstate in the $z$-direction on each lattice site.
Next we call such kind of initial state a natural state.

Let us consider the density matrix of a natural state in the eigenbasis of $\hat H$.
According to Refs.~[\onlinecite{wang2017theory},\onlinecite{PhysRevLett.107.040601}],
$\rho_{\alpha\beta}=\braket{\alpha|s}\braket{s|\beta}$
can be generally expressed as
\begin{equation}\label{NESSH}
\rho_{\alpha\beta}=D^{-1}\left(\bar{E}\right)\tilde\rho_{\alpha\alpha}
\delta_{\alpha\beta}+D^{-\eta}\left(\bar{E}\right)
f\left(\omega,\bar{E}\right)R^s_{\alpha\beta},
\end{equation}
where the first and second terms on the right-hand side are for the diagonal
and off-diagonal elements, respectively. We see the similarity between
Eq.~\eqref{NESSH} and Eq.~\eqref{eth}. But $\rho_{\alpha\alpha}$ contains
a factor $D^{-1}$, which is due to the fact $\sum_\alpha \rho_{\alpha\alpha}=1$
and then $\rho_{\alpha\alpha}$ must decrease exponentially with the system's size.
$\tilde\rho_{\alpha\alpha}$ is the scaling-free part of diagonal element,
whose expression will be discussed in Sec.~\ref{sec:diag}. The off-diagonal
element also decreases exponentially with the system's size, therefore,
it contains a factor $D^{-\eta}$. The value of $\eta$ will be discussed in Sec.~\ref{sec:off-diag}.
$f\left(\omega,\bar{E}\right)$ can be called the
dynamical characteristic function, which contains the crucial information
for understanding the real-time dynamics from the transient regime
to long-time steady limit. $R^s_{\alpha\beta}$ is a random number with zero mean and
unit variance.

\subsection{The diagonal elements \label{sec:diag}}

Let us consider the diagonal elements in Eq.~(\ref{NESSH}).
$\rho_{\alpha\alpha}= \braket{\alpha|s}\braket{s|\alpha}$ is the probability of
finding the system in the eigenstate $\ket{\alpha}$.
Previous numerical studies~\cite{PhysRevLett.107.040601} showed that $\tilde \rho_{\alpha\alpha}$
is indeed a Gaussian function of energy
with the variance and mean determined by the initial state~\cite{PhysRevE.56.5144}.
A proof based on quantum central limit theorem~\cite{Hartmann2004} was also provided.
We find that $\tilde \rho_{\alpha\alpha}$ displays a fluctuation in addition to its Gaussian shape. Therefore, we express
$\tilde\rho_{\alpha\alpha}$ as
\begin{equation}\label{NESSHd}
\tilde\rho_{\alpha\alpha}= \rho\left(\bar{E}\right) + C_sR^s_{\alpha\alpha},
\end{equation}
where
\begin{equation}\label{eq:gauss}
\rho\left(\bar{E}\right)= \frac{1}{\sqrt{2\pi\sigma^2_s}}
e^{-\frac{(\bar{E}-\mu_s)^2}{2\sigma^2_s}}
\end{equation}
is a Gaussian function of mean $\mu_s$ and variance $\sigma_s^2$.
$C_sR^s_{\alpha\alpha}$ describes the deviation from Gaussian function.
$R^s_{\alpha\alpha}$ is a random number with zero mean and unit variance.

There is an easy way of evaluating the parameters $\mu_s$ and $\sigma^2_s$.
It is not difficult to find
\begin{equation}\label{mean}
\begin{split}
\mu_s =&\int\bar{E}\rho\left(\bar{E}\right)\mathrm{d}\bar{E} \\
=&\sum_{\alpha}E_{\alpha}\braket{\alpha|s}\braket{s|\alpha} \\ =& \bra{s} \hat H \ket{s},
\end{split}
\end{equation}
where we have used $\int\mathrm{d}E_{\alpha}D(E_{\alpha})= \sum_{\alpha}$ and
$E_{\alpha}=\bar{E}$ for diagonal elements.
Note that $\int C_sR^s_{\alpha\alpha} \bar{E}\mathrm{d}\bar{E}$ is zero
in average because $R^s_{\alpha\alpha}$ is a random number with zero mean. Similarly, we find
\begin{eqnarray}\label{var}
\sigma_s^2&=&\int\bar{E}^2\rho\left(\bar{E}\right)\mathrm{d}
\bar{E}-\left(\int\bar{E}\rho\left(\bar{E}\right)\mathrm{d}\bar{E}\right)^2\nonumber\\
&=&\bra{s}\hat{H}^2\ket{s}-\bra{s}\hat{H}\ket{s}^2   \nonumber\\
&=&\sum_{s'\ne s}H^2_{ss'}.
\end{eqnarray}
Here the sum with respect to $s'$ is over the natural states which form a basis of the Hilbert space.
Different from $\mu_s$, $\sigma_s^2$ is determined by
the off-diagonal elements of Hamiltonian in the natural basis.

\subsection{The off-diagonal elements \label{sec:off-diag}}

Next we consider the off-diagonal elements in Eq.~(\ref{NESSH}).
Since $\ket{s}$ is a pure state, we have $\text{Tr}\left[\hat\rho^2\right]=
\text{Tr}\left[\hat\rho\right]=1$ and then $\displaystyle\sum_{\alpha,\beta}\left|\rho_{\alpha,\beta}\right|^2=1$.
Therefore, we expect $\rho_{\alpha,\beta}$ to scale as $D^{-1}$, which indicates $\eta=1$.
This is different from what is assumed in Ref.~[\onlinecite{wang2017theory}] ($\eta=3/2$).
Our numerical results support $\eta=1$ rather than $\eta=3/2$.

RMT tells us that the eigenstates are random vectors in the Hilbert space.
$\braket{s|\alpha}$ and $\braket{s|\beta}$ are then random numbers, so
is $\rho_{\alpha\beta}$. In Eq.~\eqref{NESSH}, the randomness of $\rho_{\alpha\beta}$
manifests itself in the factor $R^s_{\alpha\beta}$ which by definition
has zero mean and unit variance. The variance of $\rho_{\alpha\beta}$ is not a constant,
but depends on the energies $E_\alpha$ and $E_\beta$. This dependence
is given by the dynamical characteristic function
$f\left(\omega,\bar{E}\right)$ which is a function of $\bar{E}=(E_{\alpha}
+E_{\beta})/2$ and $\omega=E_{\alpha}-E_{\beta}$.

The shape of $f\left(\omega,\bar{E}\right)$ is connected to the diagonal
elements of density matrix. Because $\hat{\rho}=\ket{s}\bra{s}$ is a pure state,
we have $\left|\rho_{\alpha\beta}\right|^2=\rho_{\alpha\alpha}\rho_{\beta\beta}$.
The mean of $\left|\rho_{\alpha\beta}\right|^2$ is $D^{-2}\left|f\right|^2$ according to Eq.~(\ref{NESSHd}).
At the same time, the mean of $\rho_{\alpha\alpha}\rho_{\beta\beta}$ can be obtained
by using Eq.~(\ref{NESSHd}) and~\eqref{eq:gauss}. We then find
\begin{equation}\label{rhoab}
\begin{split}
f(\omega,\bar{E})=& \frac{D^2(\bar{E})}{D(\bar{E}+\omega/2) D(\bar{E}-\omega/2)}
\bigg(C_s^2 \overline{R^s_{\alpha\alpha}R^s_{\beta\beta}}  \\  & + \frac{e^{-\left(\left(\bar{E}+\omega/2-\mu_s\right)^2+
\left(\bar{E}-\omega/2-\mu_s\right)^2\right)/\left(2\sigma_s^2\right)}}{2\pi\sigma_s^2} \bigg),
\end{split}
\end{equation}
where $\overline{R^s_{\alpha\alpha}R^s_{\beta\beta}}$ is the mean of
${R^s_{\alpha\alpha}R^s_{\beta\beta}}$. The correlation between $R^s_{\alpha\alpha}$
and $R^s_{\beta\beta}$ may be nonzero~\cite{PhysRevE.99.042139}, thereafter,
$f(\omega,\bar{E})$ may be different from a Gaussian function.

By substituting Eqs.~(\ref{eth}) and~(\ref{NESSH}) into Eq.~(\ref{quench}), one can obtain
a formula for $O(t)$~\cite{wang2017theory}. In this paper, we do not consider the
observables but focus on testing the assumptions about density matrix.

\section{\label{sec:level 3}Diagonal elements in the spin lattice models}

We test the assumptions in the spin lattice models.
We consider the two-dimensional transverse field Ising model
(2D-TFIM) and one-dimensional disordered XXZ model.
The Hamiltonian of 2D-TFIM is
\begin{equation}\label{isingh}
\hat{H}_{Ising}=-J\sum_{\langle i,j\rangle}\hat{\sigma}^z_i\hat{\sigma}^z_j+g\sum_i\hat{\sigma}^x_i,
\end{equation}
where $\langle i,j\rangle$ denotes a pair of nearest-neighbor sites.
$\hat{\sigma}^z_i$ and $\hat{\sigma}^x_i$ are the Pauli matrices on site $i$.
The ferromagnetic coupling $J$ is set to the energy unit and $g$
denotes the transverse field. The total number of sites in numerical simulation
is denoted by $N$. This model was found to display the signatures of quantum
chaos~\cite{mondaini2017eigenstate,mondaini2016eigenstate}.
We choose the natural state $\ket{s}$ to be the common eigenstate of
$\{\hat \sigma_i^z\}$. By using Eq.~\eqref{var}, we easily find
$\sigma^2_s=Ng^2$. The fluctuation of energy density in state $\ket{s}$
is then $\frac{\sigma^2_s}{N^2}=\frac{g^2}{N}$, which vanishes
in the thermodynamic limit $N\to\infty$.

The second model we studied is the one-dimensional XXZ model:
\begin{equation}\label{xxzm}
\hat{H}_{XXZ}=-J\sum_i\left(\hat{\sigma}_i^x\hat{\sigma}_{i+1}^x
+\hat{\sigma}_i^y\hat{\sigma}_{i+1}^y+\hat{\sigma}_i^z
\hat{\sigma}_{i+1}^z\right)+\sum_ih_i\hat{\sigma}_i^z,
\end{equation}
where $h_i\in\left[-h, h\right]$ is a random number with uniform distribution
and $h$ denotes the disorder strength. Again $J$ is set to unity.
The XXZ model without disorder is integrable, but the disorder
destroys integrability. The XXZ model is in the many-body localized phase
in the case of strong disorder. In our study, we choose a small $h$
for avoiding localization.

\begin{figure}[htbp]
	\centering
	\includegraphics[width=0.45\textwidth]{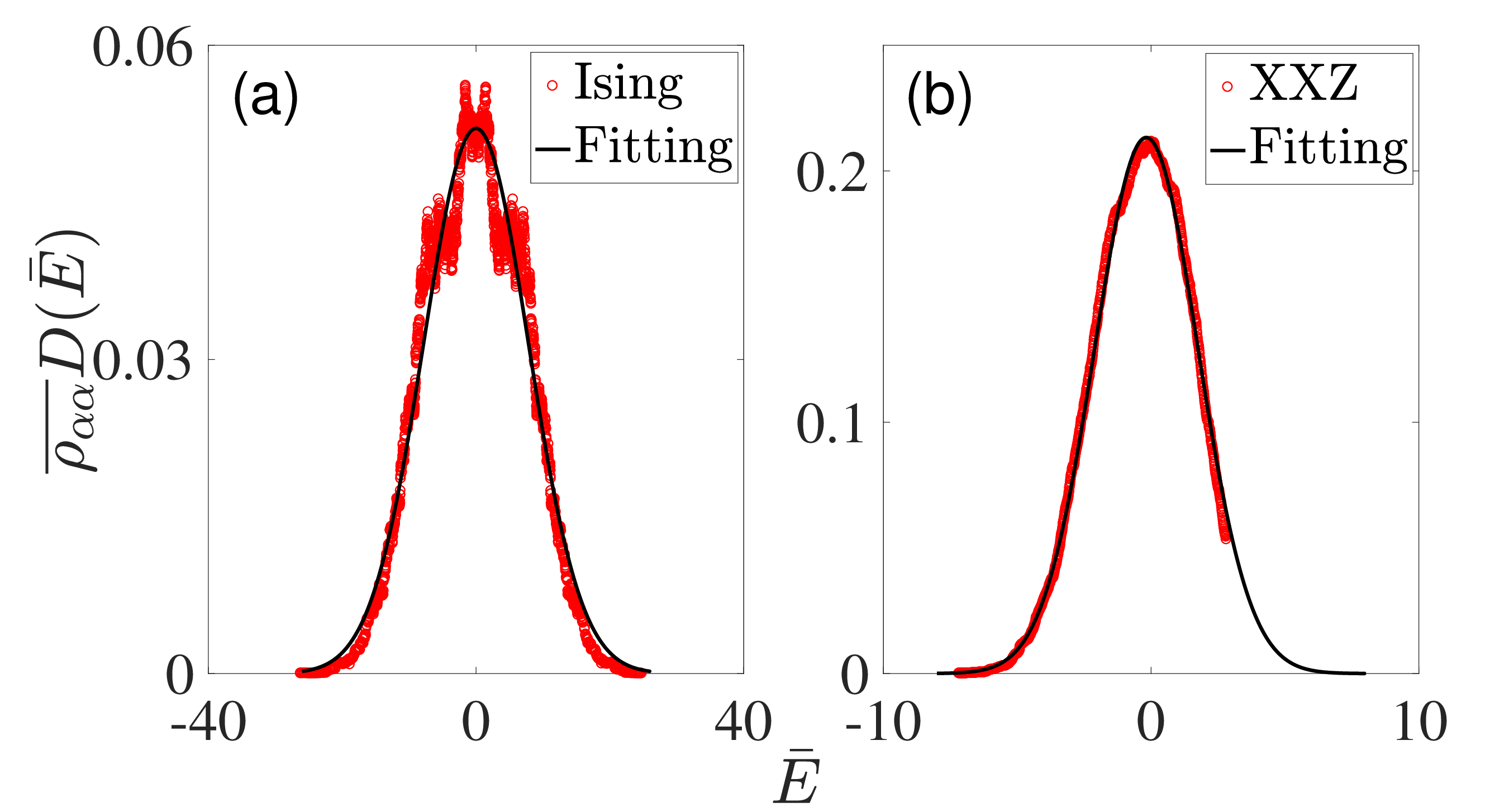}
	\caption{(Color online) $\overline{\rho_{\alpha\alpha}}
	D$ as a function of $\bar{E}$ for (a) 2D-TFIM with $g=2$,
	$\Delta\bar{E}=1$ and $N=12$, and (b) disordered XXZ chain
	with $h=0.05$, $\Delta\bar{E}=1$ and $N=16$. The black solid lines are Gaussian
	functions with $\mu_s=4.58\times10^{-6}$ and $\sigma^2_s=61.3$ for (a) and
	$\mu_s=-0.18$ and $\sigma^2_s=3.73$ for (b).}\label{fig1}
\end{figure}
In order to obtain the density matrix, we diagonalize the model Hamiltonians.
Both models have some symmetries so that their Hamiltonians are block diagonal matrices.
For the 2D-TFIM, we choose a lattice of specific shape that
breaks the geometric symmetries (see Ref.~[\onlinecite{mondaini2016eigenstate}]).
For the XXZ model, we follow Ref.~[\onlinecite{bertrand2016anomalous}] and consider
the subspace of Hilbert space with $\sigma^z=0$ where $\sigma^z$ is the eigenvalue
of $\hat{\sigma}^z=\sum_i\hat{\sigma}_i^z$.

\begin{figure}[htbp]
	\centering
	\includegraphics[width=0.45\textwidth]{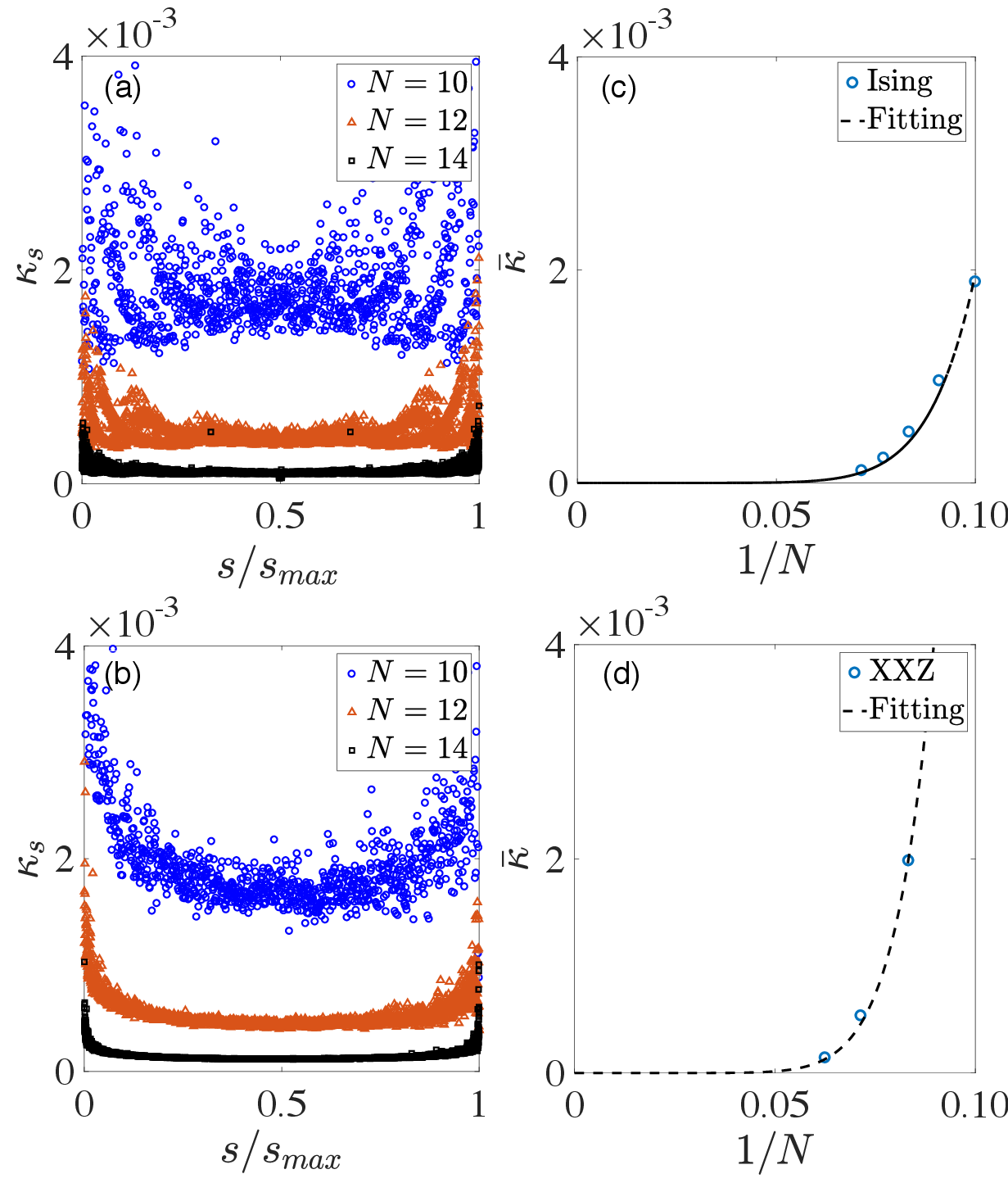}
	\caption{(Color online) Panels~(a) and~(b) plot $\kappa_s$ for the 2D-TFIM with $g=2$
	and XXZ model with $h=0.05$, respectively.
	The natural states are sorted, and $s$ denotes
	their sequence number and $s_{max}$ denotes the total number of natural states
	(the dimension of Hilbert space).
	Different dot types and colors are for different $N$.
	Panels (c) and (d) plot $\bar{\kappa}$ as a function of $1/N$ in
	the 2D-TFIM and XXZ model, respectively. The dashed lines
	are the fitting functions $\bar{\kappa} \propto \left(1/N\right)^z$.}\label{fig2}
\end{figure}
According to Eq.~(\ref{NESSHd}), the diagonal element $\rho_{\alpha\alpha}$
is a Gaussian function in average. We use $\overline{\rho_{\alpha\alpha}}$ to denote
the average of $\rho_{\alpha\alpha}$ over a thin energy shell centered at $\bar{E}$.
The width of energy shell is set to $2\Delta\bar{E}=2$ so that it contains
enough number of eigenstates.
It is worth mentioning that $\Delta\bar{E}$ can be made smaller
and smaller as the system's size increases.
And in thermodynamic limit, $\Delta\bar{E} $ can be made arbitrarily small
while there are still infinite number of states in the shell.
Fig.~\ref{fig1} shows $\overline{\rho_{\alpha\alpha}} D$ as a function of $\bar{E}$.
It does fit a Gaussian function (the solid line).

$\rho_{\alpha\alpha}$ fluctuates around its average $\overline{\rho_{\alpha\alpha}}$.
We define the amplitude of fluctuation to be
\begin{equation}\label{eq:kappadef}
\kappa_s= \sqrt{\displaystyle \overline{\left(\rho_{\alpha\alpha}- \overline{\rho_{\alpha\alpha}}\right)^2} },
\end{equation}
where the overline denotes the average over $\alpha$.
Fig.~\ref{fig2}(a) and~(b) plot $\kappa_s$ for different natural states.
It is clear that $\kappa_s$ has a concentrated distribution.
As the system's size increases from $N=10$ to $N=14$,
the spread of $\kappa_s$ between different $s$ decreases.
We guess that $\kappa_s$ should be independent of the initial state
$\ket{s}$ for sufficiently large $N$. We study the average of $\kappa_s$
over ${s}$, which is denoted by $\bar{\kappa}$.
Fig.~\ref{fig2}(c) and~(d) display how $\bar{\kappa}$ changes with the system's size.
For both 2D-TFIM and XXZ model, $\bar{\kappa}$ decreases with
increasing $N$ and vanishes in the thermodynamic limit $N\to\infty$.

\section{\label{sec:level 4}Off-diagonal elements in the spin lattice models}

\begin{figure}[htbp]
	\centering
	\includegraphics[width=0.45\textwidth]{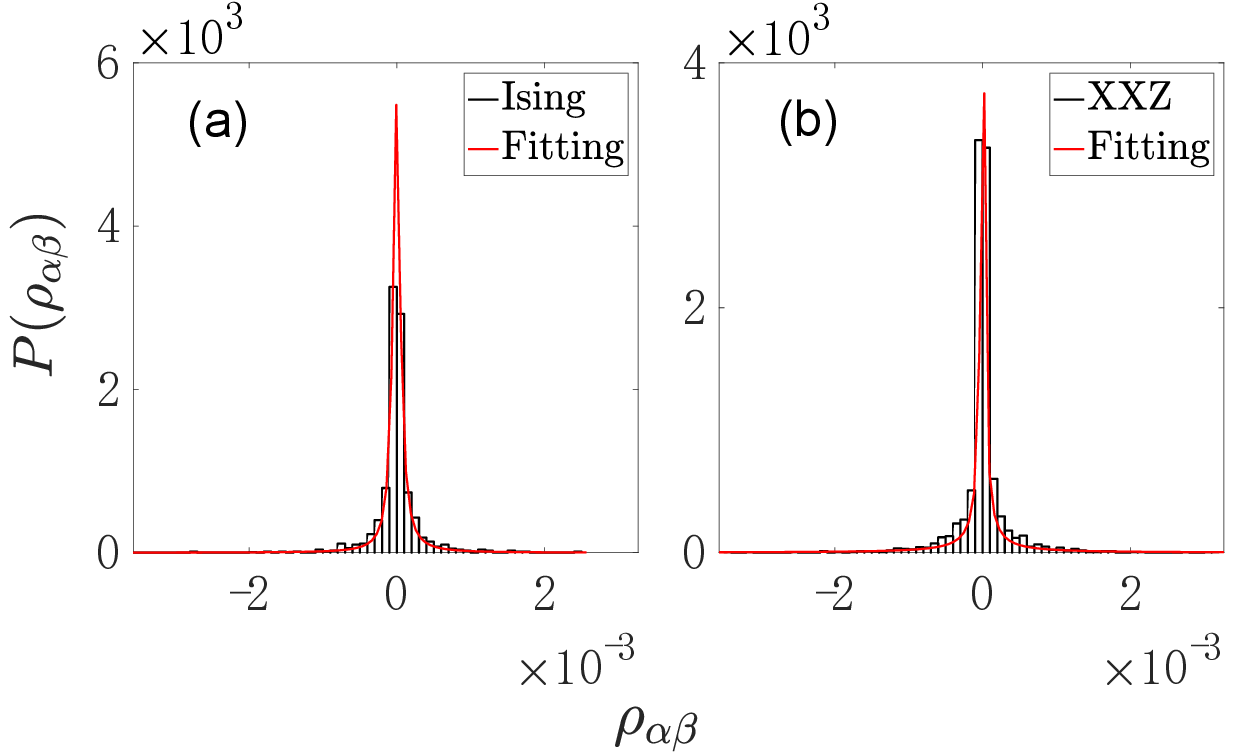}
	\caption{(Color online) The distribution of $\rho_{\alpha\beta}$
	for (a) 2D-TFIM at $g=2$ and $N=12$ and (b) XXZ model at $h=0.05$ and $N=14$.
	The energy box is centered at $\bar{E}=0$ and $\omega=1$ with the
	sides $\Delta{\bar{E}}=0.1$ and $\Delta\omega=0.1$.
	The red lines are the stable distributions with the parameters
	$a=0.99$, $b=0.05$, $c=5.78\times 10^{-5}$ and $\delta=-2.37\times10^{-6}$ for panel~(a),
	and $a=0.51$, $b=-5.51\times10^{-4}$, $c=1.67\times 10^{-5}$ and $\delta=-1.96\times10^{-9}$
	for panel~(b).}\label{fig3}
\end{figure}

Next we study $\rho_{\alpha\beta}$ for $\alpha\neq\beta$. According to Eq.~\eqref{NESSH}, $\rho_{\alpha\beta}$
is a random number. We consider the set of $\rho_{\alpha\beta}$ within a small rectangular energy box centered at
$(\bar{E},\omega)$ with the sides $2\Delta{\bar{E}}$ and $2\Delta{\omega}$.
In other words, $\alpha$ and $\beta$ satisfy $\bar{E}-\Delta{\bar{E}}<(E_\alpha +E_\beta)/2 <
\bar{E}+\Delta{\bar{E}}$ and $\omega-\Delta{\omega}<E_\alpha -E_\beta<
\omega+\Delta{\omega}$. We then obtain the distribution of $\rho_{\alpha\beta}$ which
is plotted in Fig.~$\ref{fig3}$.
It is clear that the distribution is symmetric with respect to zero, indicating
that the mean of $\rho_{\alpha\beta}$ is zero. And the distribution
functions have a similar shape for the TFIM and XXZ models. They are also
similar to the distribution function for the fermionic model studied previously~\cite{wang2017theory}.
Our finding suggests that the distribution of $\rho_{\alpha\beta}$ may be independent
of the model.

We fit the histogram of $\rho_{\alpha\beta}$ to the stable distribution (the red lines in Fig.~$\ref{fig3}$),
whose probability density is defined as
\begin{equation}
\begin{split}
P\left( x \right)=& \frac{1}{2\pi} \int dp \,
e^{-ip(x-\delta)} \\ & \times e^{-c^a \left|p \right|^a  \left[1+i b\, \text{sign}\left(p\right) \tan(\pi a/2) 
\left( \left(c\left|p \right| \right)^{1-a} - 1\right)\right]},
\end{split}
\end{equation}
where $a$, $b$, $c$ and $\delta$ are the parameters. For the 2D-TFIM,
we obtain $a=0.99$, $b=0.05$, $c=5.78\times 10^{-5}$ and $\delta=-2.37\times10^{-6}$.
For the XXZ chain, we obtain $a=0.51$, $b=-5.51\times10^{-4}$,
$c=1.67\times 10^{-5}$ and $\delta=-1.96\times10^{-9}$.
$\delta$ is the location parameter, which is almost zero for both models,
indicating that the distribution is symmetric to zero.
$c$ is the scale parameter, which is also small. The shape parameters $a$ and $b$ measure
the concentration and asymmetry of the distribution, respectively.

\begin{figure}[htbp]
	\includegraphics[width=0.45\textwidth]{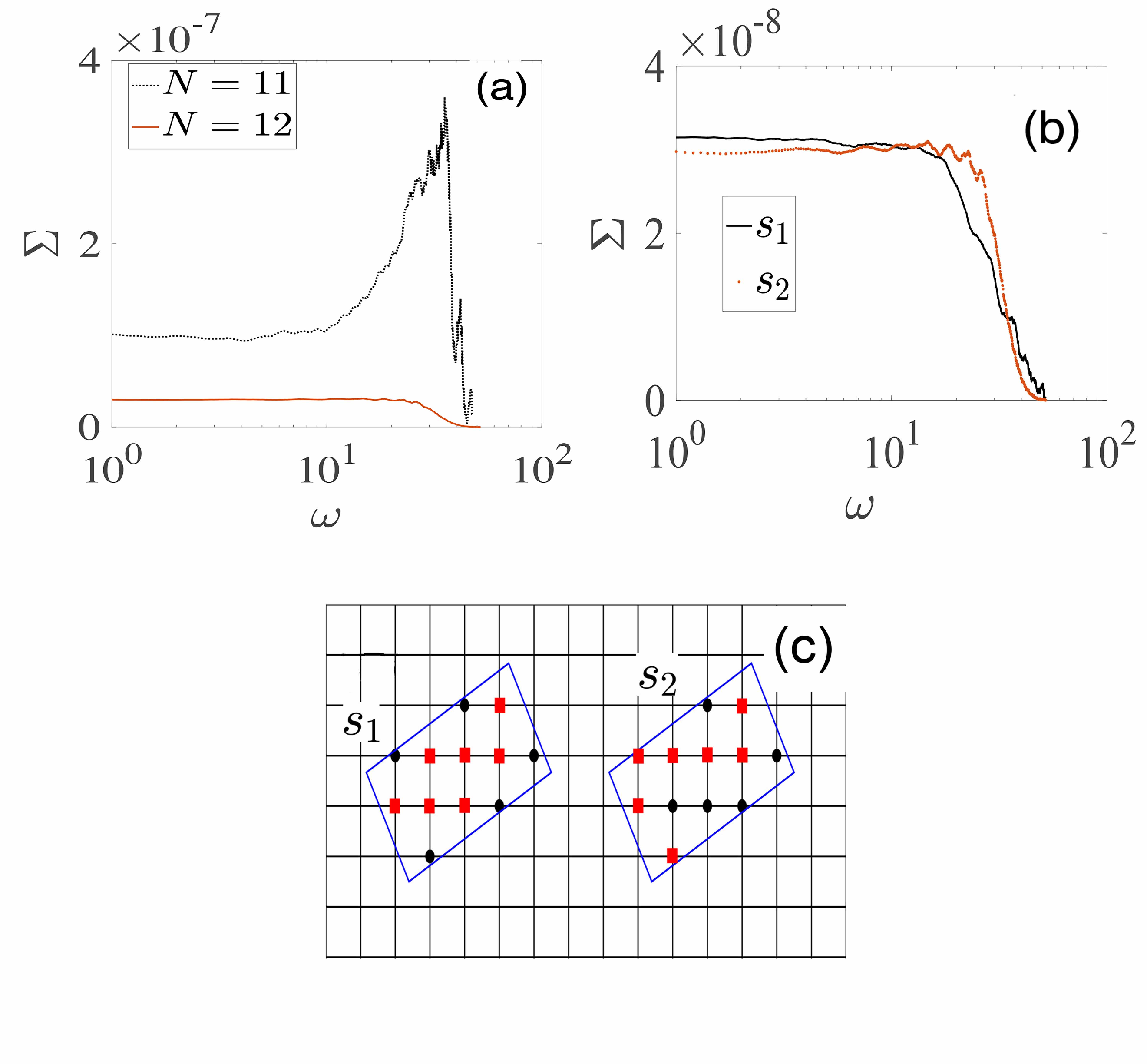}
	\caption{(Color online) The variance of $\rho_{\alpha\beta}$ as a function
	of $\omega$ is plotted for the 2D-TFIM.
	Panel (a) shows $\Sigma(\omega)$ at different system's size.
	Panel (b) shows $\Sigma(\omega)$ for two
	different initial states $\ket{s_1}$ and $\ket{s_2}$ whose
	spin configurations are displayed in panel~(c) with
	the circles and squares representing the spin-up and spin-down states, respectively.}\label{fig4}
\end{figure}

We numerically calculate the variance of $\rho_{\alpha\beta}$ (denoted as
$\Sigma(\bar{E},\omega)$) by averaging $\left|\rho_{\alpha\beta}\right|^2$ over
each energy box. We then average $\Sigma(\bar{E},\omega)$ over $\bar{E}$ and obtain a function $\Sigma(\omega)$,
which shows how the dynamical characteristic function changes
with $\omega$. The results are plotted in Fig.~\ref{fig4}. $\Sigma(\omega)$ is a smooth
function of $\omega$, as we expect. And it displays a plateau at small $\omega$.
For large $\omega$, $\Sigma(\omega)$ decays exponentially to zero.
This is believed to be a typical feature of it.
For $N=11$, $\Sigma(\omega)$ displays a peak at $\omega>10$
(see Fig.~\ref{fig4}(a)), possibly because $N=11$ is too small.
This peak vanishes as we choose $N=12$.

The dynamical characteristic function depends on
the initial state. Fig.~\ref{fig4}(b) shows $\Sigma(\omega)$
for two different states $\ket{s_1}$ and $\ket{s_2}$, while the spin configurations
of $s_1$ and $s_2$ are shown in Fig.~\ref{fig4}(c).
$\Sigma(\omega)$ of $s_1$ and $s_2$ differ from each other,
but the difference is not significant. They both show a plateau at small $\omega$
and an exponential decay at large $\omega$.

\begin{figure}[htbp]
	\centering
	\includegraphics[width=0.45\textwidth]{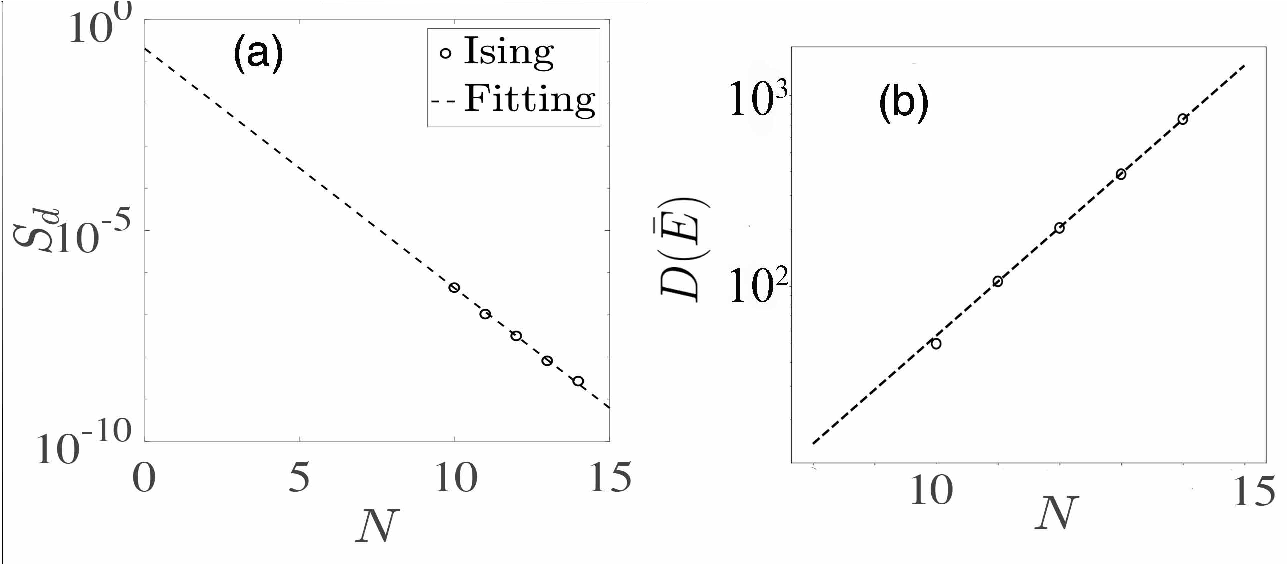}
	\caption{(Color online) 
	(a) The plateau of $\Sigma(\omega)$ at small $\omega$ is plotted 
	as a function of the system's size. The dotted line is $0.21 e^{-1.31 N}$. 
	(b) $D\left(\bar{E}\right)$ at $\bar{E}=0$ is plotted as a 
	function of $N$. The dotted line is $0.081e^{0.65N}$.}\label{fig5}
\end{figure}

According to Eq.~\eqref{NESSH}, $\rho_{\alpha\beta}$
should be a random number of zero mean and the variance $D^{-2}f^2$.
$\Sigma$ should then decrease exponentially as the system's size increases.
In Fig.~\ref{fig4}(a), we already see that the value of $\Sigma$
decreases with increasing $N$.
We use $S_d$ to denote the value of $\Sigma(\omega)$ at the small-$\omega$ plateau.
Fig.~\ref{fig5}(a) displays $S_d$ as a function of $N$.
We clearly see that $S_d$ decays exponentially with increasing $N$.
We also plot the density of states for different $N$
in Fig.~\ref{fig5}(b). Here we choose $\bar{E}=0$, i.e. at the middle of the energy band,
and then obtain $D(\bar{E})$ by counting all the levels $\bar{E}\in(-0.5, 0.5)$. By curve fitting,
we find that $\Sigma$ scales as $ \displaystyle e^{-1.31N}$ while $D(\bar{E})$
scales as $\displaystyle e^{0.65N}$. Therefore, we approximately
have $\Sigma \sim D^{-2}(\bar{E})$, which indicates that 
$\rho_{\alpha\beta}$ contains a factor  $D^{-1}(\bar{E})$.
This result is different from that in Ref.~[\onlinecite{wang2017theory}],
but coincides with the fact $\text{Tr} \left[ \hat\rho^2\right]=1$.

\section{\label{sec:level 6}summary}

In summary, the numerics support that the density matrix in the spin lattice models
can be expressed as Eq.~\eqref{NESSH}.
The diagonal element $\rho_{\alpha\alpha}$ fluctuates
around a Gaussian function with the fluctuation vanishing in thermodynamic limit.
The off-diagonal elements $\rho_{\alpha\beta}$ are random numbers which have
the stable distribution. The variance of $\rho_{\alpha\beta}$, i.e. the squared dynamical characteristic function,
exhibits a plateau at low frequencies but an exponential decay at high frequencies.

\section*{acknowledgements}
This work is supported by NSF of China under Grant Nos.~11774315 and~11304280.
Pei Wang is also supported by the Junior Associates programm
of the Abdus Salam International Center for Theoretical Physics.

\end{document}